\documentclass[unsortedaddress, aps, prd, reprint, nofootinbib]{revtex4-2}

\usepackage{graphicx}
\usepackage{amsmath,amssymb,mathtools}
\usepackage{dcolumn}
\usepackage{bm}
\usepackage{xcolor}
\usepackage{soul}
\usepackage{braket}
\usepackage[separate-uncertainty=true,multi-part-units=single]{siunitx}
\usepackage{scalerel}

\definecolor{Mathematica1}{rgb}{0.368417, 0.506779, 0.709798}
\definecolor{Mathematica2}{rgb}{0.880722, 0.611041, 0.142051}
\definecolor{Mathematica3}{rgb}{0.560181, 0.691569, 0.194885}
\definecolor{darkred}{rgb}{0.545,0,0}
\definecolor{dullblue}{rgb}{0,0.298,0.49}
\definecolor{blue3}{RGB}{31, 119, 180}
\usepackage[colorlinks=true
,urlcolor=dullblue
,anchorcolor=blue3
,citecolor=dullblue
,filecolor=blue3
,linkcolor=darkred
,menucolor=blue3
,pagecolor=blue3
,linktocpage=true
,pdfproducer=medialab
]{hyperref}

\newcommand{\dd}{\mathop{\mathrm{d}\!}{}}

\newcommand{\slab}[1]{{\textsc{#1}}}

\newcommand{\sminus}{{\protect \scalebox {0.7}[0.7]{$-$}}}

\renewcommand{\vec}[1]{\boldsymbol{\mathbf{#1}}}

\DeclarePairedDelimiter{\abs}{\lvert}{\rvert}

\def\beq{\begin{equation}}
\def\eeq{\end{equation}}

\begin{document}

\title{Scattering of wave dark matter by supermassive black holes}

\author{Giovanni Maria Tomaselli}

\affiliation{School of Natural Sciences, Institute for Advanced Study, Princeton, NJ 08540, USA}

\begin{abstract}

Recent simulations of wave dark matter around black hole binaries revealed the formation of a universal density profile that co-rotates with the binary. We derive this profile from first principles, interpreting it as the steady state of a scattering process. We find that the scattering becomes particularly efficient when the ratio of the binary separation to the dark matter's de Broglie wavelength assumes certain discrete values, which can be interpreted as bound state resonances. After estimating the amount of dark matter that undergoes this type of scattering off supermassive black hole binaries at galactic centers, we demonstrate that the process can induce an observable modification of the slope of the pulsar timing array spectrum. This opens up a new possibility to gain insights on the nature of dark matter from observations of low-frequency gravitational waves.

\end{abstract}

\maketitle

\section{Introduction}

Galactic centers are very dynamic and busy environments. They are the home of supermassive black holes (BHs), as well as the regions hosting the highest densities of baryonic and dark matter (DM). During a merger of two galaxies, the respective supermassive BHs are expected to migrate towards the center of the resulting galaxy, eventually forming a binary system. These supermassive BH binaries are considered to be the most likely source of the gravitational-wave (GW) background recently detected by pulsar timing arrays (PTAs) \cite{NANOGrav:2023gor,EPTA:2023fyk,Reardon:2023gzh,Xu:2023wog,NANOGrav:2023hfp,EPTA:2023xxk}.

It is natural to ask how these binaries interact with their dense environment, such as the DM at halo center. The fundamental nature of DM is unknown: in particular, the mass $\mu$ of its constituents is very poorly constrained, with candidates spanning a range of about 90 orders of magnitude, see e.g.~\cite{Cirelli:2024ssz} for a recent review. The case $\mu\lesssim\SI{30}{eV}$, where the average interparticle separation is smaller than the de Broglie wavelength at typical galactic densities, is known as ``wave dark matter'' \cite{Hu:2000ke,Hui:2016ltb,Hui:2021tkt}. In this scenario, DM behaves as a classical wave, rather than a collection of particles. A variety of astrophysical probes have been used to place lower bounds on $\mu$, ranging from $\SI{e-21}{eV}$ to $\SI{e-19}{eV}$ \cite{Armengaud:2017nkf,Irsic:2017yje,Kobayashi:2017jcf,Rogers:2020ltq,Schutz:2020jox,Banik:2019smi,DES:2020fxi,Dalal:2022rmp}.

A series of recent studies used numerical relativity simulations to understand the interaction of wave DM with a BH binary \cite{Bamber:2022pbs,Aurrekoetxea:2023jwk,Aurrekoetxea:2024cqd}. The main finding of these works is that, regardless of the initial conditions imposed on the DM density field, a universal quasi-stationary DM configuration forms around the BH binary. Depending on the overall density normalization, this DM structure can affect the BH dynamics by accelerating their inspiral, as well as lead to a richer phenomenology when DM self-interactions are included.

The goal of this paper is twofold. First, inspired by the results of \cite{Bamber:2022pbs,Aurrekoetxea:2023jwk,Aurrekoetxea:2024cqd}, we study the stationary DM structure from first principles, without the use of simulations. Such a semi-analytical description uncovers the underlying physical mechanism, facilitates better interpretation of the results, and allows us to explore the parameter space easily, which leads to the discovery of qualitatively new features. The second goal is to determine in which scenarios the BH-DM interaction leads to observable effects, focusing on the spectrum of the stochastic GW background detected by PTAs.

\section{Scattering of dark matter}
\label{sec:scattering}

\subsection{Setup}

Neglecting self-interactions, wave DM is described by a real scalar field $\Phi$, obeying the Klein-Gordon equation
\beq
\bigl(\Box-\mu^2\bigr)\Phi=0\,,
\label{eqn:kg}
\eeq
where we used natural units, $\hbar=c=1$. The limit $\mu\to\infty$ recovers the physics of point particles. In a weak gravitational potential and for small velocities, the ansatz $\Phi=\psi e^{-i\mu t}/\sqrt{2\mu}+\text{c.c.}$ reduces equation \eqref{eqn:kg} to a Schrödinger-like form,
\beq
\biggl(-\frac{\nabla^2}{2\mu}+V\biggr)\psi=i\frac{\partial\psi}{\partial t}\,,
\label{eqn:schrodinger}
\eeq
where $V$ is the gravitational potential. In principle, $V$ features a contribution from the DM energy density, which results in a Gross-Pitaevskii interaction and makes \eqref{eqn:schrodinger} nonlinear. Close enough to a BH binary, however, the gravity sourced by DM will be negligible compared to that of the BHs. For two BHs of equal masses $M/2$ on circular orbits of radius $R$, we can thus write
\beq
V(\vec r,t)=-\frac{GM\mu}2\biggl(\frac1{\abs{\vec r-\vec R(t)}}+\frac1{\abs{\vec r+\vec R(t)}}\biggr)\,,
\label{eqn:V}
\eeq
where, in spherical coordinates $(r,\theta,\phi)$ with the $z$ axis aligned with the binary's angular momentum, we have $\vec R(t)=(R,\pi/2,\Omega t)$ and $\Omega^2=GM/(8R^3)$. We treat BHs as point particles: in the problem studied in this paper, accretion of DM is negligible when the BH's sizes are much smaller than their separation.

The time-dependent Schrödinger equation given by \eqref{eqn:schrodinger} and \eqref{eqn:V} has been studied in \cite{Ikeda:2020xvt}. This work showed the existence of quasi-bound states of the scalar field around the binary, which share many similarities with the 1-electron states of a diatomic molecule in the limit where $\Omega$ is small, i.e., when the time-dependence is neglected. The scenario of interest for the present work also allows to eliminate the time dependence, but in a different way.

First of all, it is convenient to expand the wavefunction in spherical harmonics as
\beq
\psi(\vec r,t)=\sum_{\omega\ell m}\frac{U_{\ell m}(r)}rY_{\ell m}(\theta,\phi)e^{-i\omega t}\,.
\label{eqn:psi-decomposition}
\eeq
The universal quasi-stationary configuration observed in \cite{Bamber:2022pbs,Aurrekoetxea:2023jwk} appears to rigidly rotate with the same angular frequency as the binary. Hence, let us look for a solution where the density $\rho\propto\abs{\psi}^2$ is not a function of $\phi$ and $t$ separately, but only of the combination $\phi-\Omega t$. The DM density gets contributions from terms of the form
\beq
\abs{\psi}^2\supset Y_{\ell'm'}^*e^{i\omega't}Y_{\ell m}e^{-i\omega t}\propto e^{-i(\omega-\omega')t+i(m-m')\phi}\,.
\eeq
We thus impose $\omega-\omega'=\Omega(m-m')$, or equivalently
\beq
\omega=\omega_0+m\Omega\equiv\omega_m\,,
\label{eqn:omega_m}
\eeq
where $\omega_0$ is the frequency of modes with $m=0$.

Expanding $V$ in multipoles, equation \eqref{eqn:schrodinger} then becomes
\beq
\begin{split}
\frac{\dd^2U_{\ell m}}{\dd r^2}={}&{-k_m^2U_{\ell m}}+\frac{\ell(\ell+1)}{r^2}U_{\ell m}\\
&-GM\mu^2\sum_{\ell_*}B^{\ell'm'}_{\ell m;\ell_*}\mathcal F_{\ell_*}(r)U_{\ell'm'}\,,
\end{split}
\label{eqn:d2U}
\eeq
where $k_m^2=2\mu\omega_m$ and we defined
\beq
\mathcal F_{\ell_*}(r)=\frac{r^{\ell_*}}{R^{\ell_*+1}}\Theta(R-r)+\frac{R^{\ell_*}}{r^{\ell_*+1}}\Theta(r-R)
\eeq
and
\beq
\begin{split}
B^{\ell'm'}_{\ell m;\ell_*}={}&((-1)^{m'}+(-1)^m)\sqrt{\frac{4\pi(2\ell+1)(2\ell'+1)}{(2\ell_*+1)}}\\
{}&\times\begin{pmatrix}\ell_* & \ell' & \ell\\ 0& 0 & 0\end{pmatrix}\begin{pmatrix}\ell_* & \ell' & \ell\\ -m_* & m' & -m\end{pmatrix}Y_{\ell_*,m_*}\biggl(\frac\pi2,0\biggr)\,,
\end{split}
\eeq
where $m_*=m'-m$.

At first sight, the equations seem to depend on two independent dimensionless parameters, namely $GM\mu$ and $\mu R$. However, when the frequency is fixed according to formula \eqref{eqn:omega_m}, close inspection of \eqref{eqn:d2U} reveals that it actually only depends on the product of those two parameters, i.e.\ on the combination
\beq
\beta\equiv GM\mu^2R\,.
\eeq
The physical meaning of $\beta$ can be understood by writing $\beta=(R/\lambda_{\text{dB}})^2$, where $\lambda_{\text{dB}}=(\mu v_0)^{-1}$ is the typical de Broglie wavelength of the scalar field and $v_0=\sqrt{GM/R}$. The combination $\omega_0R$ is also dimensionless, but it is related to the boundary conditions rather than the physics of the problem, as will be discussed next.

\subsection{Boundary conditions}

Apart from equation \eqref{eqn:omega_m}, we need one additional physical assumption to find the stationary configuration: an appropriate choice of boundary conditions. The initial data used in \cite{Bamber:2022pbs,Aurrekoetxea:2023jwk} specify the value of $\Phi$ in a simulation box, for example by setting it to a constant throughout space, which is then evolved over time. Replicating analytically such an initial condition would lead to cumbersome and uninformative expressions.

We setup instead a scattering problem. Generally speaking, one of the terms appearing in \eqref{eqn:psi-decomposition} will be allowed to have an ingoing component at $r\to\infty$, with a certain given amplitude. The amplitude of the outgoing components of all modes will then be computed by solving equation \eqref{eqn:d2U}. These boundary conditions lead to a truly stationary configuration, where the DM density does not change secularly because DM waves are scattered away by the binary at the same rate as they are falling towards it. Note that this contrasts with the findings of \cite{Bamber:2022pbs}, where the DM profile ``grows
over time homogeneously, fed by the asymptotic reservoir of
dark matter imposed at the boundaries.'' We elaborate on this point in Appendix~\ref{app:stationary}.

We choose an isotropic DM distribution that falls in with asymptotically zero velocity. Hence, the $(\ell,m)=(0,0)$ mode will be the one containing the ingoing waves, and we set the parameter $\omega_0$ to zero, as it equals the energy of the infalling waves. By solving \eqref{eqn:d2U} at leading order for $r\to\infty$, we thus impose
\beq
U_{\ell m}\to\begin{cases}
x^{1/4}\bigl(e^{-2i\sqrt{x}}+C_{00}e^{2i\sqrt{x}}\bigr)\,, & \ell=m=0\,,\\
C_{\ell0}x^{1/4}e^{2i\sqrt{x}}\,, & \ell\ne m=0\,,\\
C_{\ell m}e^{i\varphi_m}\,, & m>0\,.
\end{cases}
\label{eqn:asymptotics}
\eeq
where we defined $x=2GM\mu^2r$ and $\varphi_m=k_mr+(GM\mu^2/k_m)\log(2k_mr)$, and the coefficients $C_{\ell m}$ are to be determined. Modes with $m<0$ have instead different asymptotic behavior: from equation \eqref{eqn:omega_m}, they must have $k_m^2<0$, hence they do not propagate to infinity. We thus impose regular boundary conditions,
\beq
U_{\ell m}\to C_{\ell m}e^{-\chi_m}\,,\quad m<0\,,
\label{eqn:bound-asymptotics}
\eeq
where $\chi_m=\kappa_mr-(GM\mu^2/\kappa_m)\log(2\kappa_mr)$ and $\kappa_m=\sqrt{-k_m^2}$. These modes are gravitationally bound to the BH binary and rigidly co-rotate with it. They are not exact eigenstates of \eqref{eqn:schrodinger} because the time-varying potential mixes them with unbound ones, so we may call them \emph{quasi-bound} modes. The requirement of a stationary DM configuration forces, through \eqref{eqn:omega_m}, that part of the DM structure must live in quasi-bound modes.

At $r=0$, the binary's center of gravity, we impose regularity of the wavefunction via $U_{\ell m}\propto r^{\ell+1}$.

\subsection{Numerical integration}

Equation \eqref{eqn:d2U} can be integrated numerically, after restricting it to the modes with $\ell\le\ell_{\text{max}}$, which are $N=(\ell_{\text{max}}+1)^2$ in total, for a certain choice of $\ell_{\text{max}}$. This allows us to completely determine the DM stationary configuration and the coefficients $C_{\ell m}$. However, the existence of quasi-bound modes makes the numerical integration challenging. The reason is that their generic large-$r$ behavior is dominated by the exponentially diverging solution, which introduces numerical instabilities and makes it hard to single out the subdominant contribution chosen in \eqref{eqn:bound-asymptotics}.

A variety of methods to overcome similar problems exist in the literature, see e.g.\ \cite{Pani:2013pma}. We proceed by expanding $U_{\ell m}$ in $N$ different Frobenius series around $r\to\infty$, in each one of which all modes have $C_{\ell m}=0$, except for a given mode $(\ell_0,m_0)$. Explicitly,
\beq
U_{\ell m}=\begin{cases}
\displaystyle e^{2i\sqrt{x}}\sum_{n=0}F_{\ell m;n}^{\ell_0m_0}x^{1/4-n/2}\,, & m_0=0\\[14pt]
\displaystyle e^{i\varphi_{m_{\scaleto{0}{2.5pt}}}}\sum_{n=0}F_{\ell m;n}^{\ell_0m_0}(k_{m_0}r)^{-n}\,, & m_0>0\\[14pt]
\displaystyle e^{-\chi_{m_{\scaleto{0}{2.5pt}}}}\sum_{n=0}F_{\ell m;n}^{\ell_0m_0}(\kappa_{m_0}r)^{-n}\,, & m_0<0
\end{cases}
\label{eqn:frobenius}
\eeq
where $F_{\ell m;0}^{\ell_0m_0}=\delta_{\ell_0\ell}\delta_{m_0m}$, while $F_{\ell m;n}^{\ell_0m_0}$ for $n\ge1$ can be determined recursively by solving \eqref{eqn:d2U} order-by-order at large $r$. In practice, the numerical routine will include terms in the series up to a certain $n_{\text{max}}$. The large-$r$ behavior of the full solution of \eqref{eqn:d2U} will be a linear superposition of the series \eqref{eqn:frobenius} for all choices of $(\ell_0,m_0)$.

Equation \eqref{eqn:frobenius} can be seen as an improved version of the boundary conditions \eqref{eqn:asymptotics}--\eqref{eqn:bound-asymptotics}. The higher-order corrections, from terms with $n\ge1$, mitigate the numerical instabilities associated with bound modes by allowing the radius where boundary conditions are imposed, say $r_{\text{max}}$, to be smaller. We choose $n_{\text{max}}$ and $r_{\text{max}}$ by trial and error, using the unitarity relation
\beq
1=\sum_{\{k_m^2=0\}}\abs{C_{\ell m}}^2+\sum_{\{k_m^2>0\}}\frac{\abs{C_{\ell m}}^2k_m}{2GM\mu^2}\,,
\label{eqn:unitarity}
\eeq
which can be derived from the mass flux conservation, as a proxy for numerical stability. Typical values we use are $n_{\text{max}}=20$ and $r_{\text{max}}/R=1+22/\beta^{1/4}$.

We first numerically integrate equation \eqref{eqn:d2U} from $r=r_{\text{max}}$ to $r=R$ with ingoing boundary conditions for the $(0,0)$ mode. We then do the same for all the $N$ possible outgoing boundary conditions given in \eqref{eqn:frobenius}. Finally, we integrate from $r=0$ to $r=R$ for $N$ independent regular boundary conditions. The full solution can be constructed by concatenating a linear superposition of those for $r<R$ with a linear superposition of those for $r>R$. If the amplitude of the ingoing $(0,0)$ mode is fixed, the remaining $2N$ free coefficients can be determined requiring that $U_{\ell m}$ and $\dd U_{\ell m}/\dd r$ are continuous at $r=R$. Because of the symmetry of the problem, all coefficients $C_{\ell m}$ where either $\ell$ or $m$ is odd turn out to be identically zero.

\subsection{Power emitted}

By construction, the DM configuration is stationary, therefore the total ingoing and outgoing mass fluxes are equal, say to $\dot M_\slab{dm}$. A useful quantity to study is the total power emitted in scalar waves, given by
\beq
P=\frac{\dot M_\slab{dm}}\mu\sum_{\{k_m^2>0\}}\frac{\abs{C_{\ell m}}^2k_m}{2GM\mu^2}\times\frac{k_m^2}{2\mu}\,,
\label{eqn:power}
\eeq
where the first factor in the sum is the flux of the number current of the $(\ell,m)$ mode, while the second factor is its energy.

\begin{figure*}
\centering
\includegraphics[]{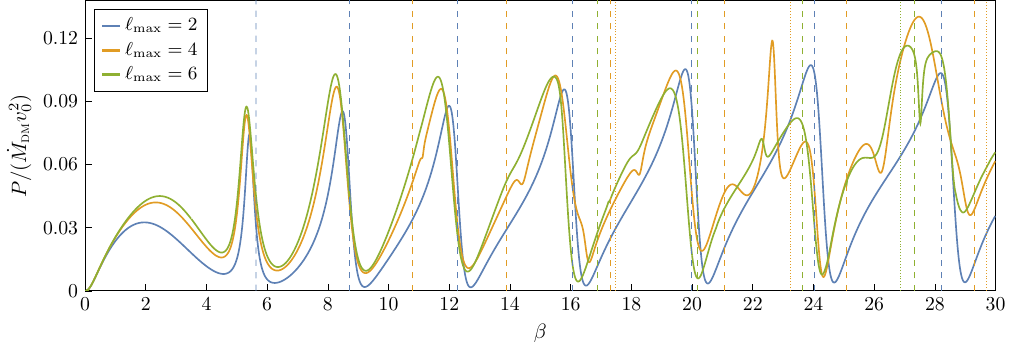}
\caption{Ratio between the power emitted in scalar waves $P$ and the infalling mass flux $\dot M_\slab{dm}$, as computed from \eqref{eqn:power} and normalized by the squared velocity $v_0^2=GM/R$, as function of $\beta$ for different values of $\ell_{\text{max}}$. The vertical lines denote the position of bound state resonances: blue dashed for $\beta_{n2\,\sminus2}$, orange dashed for $\beta_{n4\,\sminus2}$, green dashed for $\beta_{n6\,\sminus2}$, orange dotted for $\beta_{n4\,\sminus4}$, and green dotted for $\beta_{n6\,\sminus4}$.}
\label{fig:power}
\end{figure*}

\begin{figure*}
\centering
\includegraphics[]{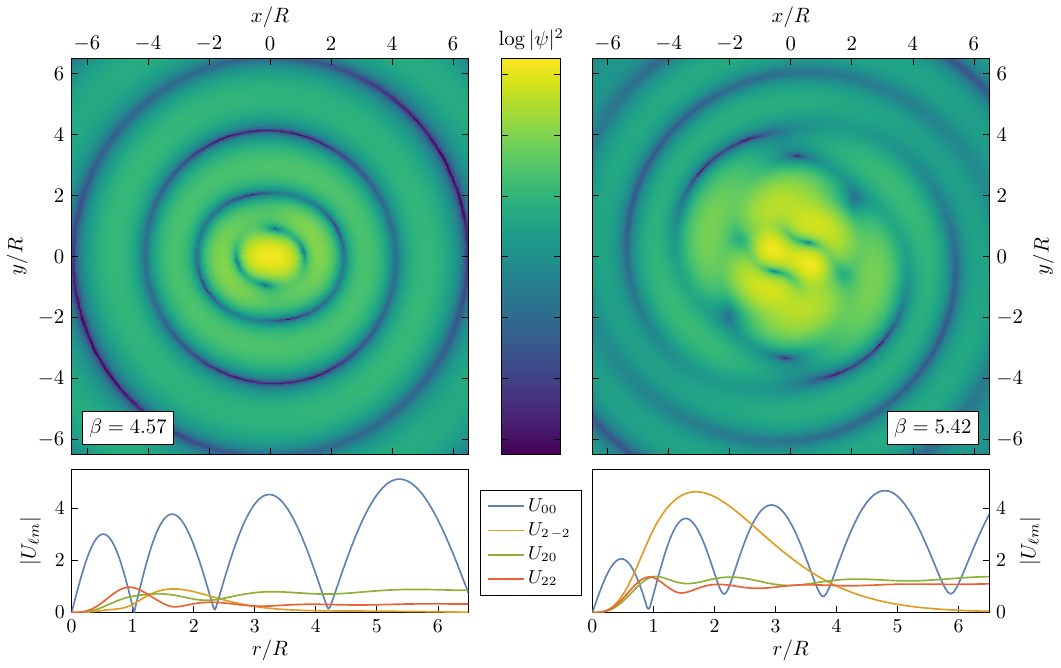}
\caption{\emph{Top panels:} DM density on the binary's plane for $\ell_{\text{max}}=2$ and two values of $\beta$: off-resonance (\emph{left}), and on resonance (\emph{right}). Off-resonance, the DM scattering is inefficient, and an approximately spherical standing wave pattern is created. On resonance, instead, a large fraction of the infalling DM is scattered into states with nonzero angular momentum, creating a spiral pattern around the binary. \emph{Bottom panels:} mode decomposition of the DM profile in the two cases described above. Off-resonance, the amplitude of the spherically symmetric $(0,0)$ mode is much larger than all the others. On resonance, the quasi-bound mode $(2,-2)$ is significantly excited. Furthermore, the mode $(2,2)$, which is the only one carrying energy to infinity, has a larger asymptotic value, leading to a higher emitted power $P$.}
\label{fig:profile}
\end{figure*}

The ratio $P/\dot M_\slab{dm}$, normalized by $v_0^2=GM/R$, is plotted in Fig.~\ref{fig:power} as a function of $\beta$, for different choices of $\ell_{\text{max}}$. The curve exhibits notable features, such as the presence of recurring peaks. As expected, $P$ vanishes for $\beta\to0$, as the ``quantum pressure'' of DM increases for small $\mu$, suppressing the effect of the BHs' gravity. When $\ell_{\text{max}}\ge4$, the curve does not show a precisely regular nor periodic behavior at large $\beta$, however the height of the peaks appears to remain roughly constant. For large $\beta$, higher values of $\ell_{\text{max}}$ would be needed for better convergence of the curves; however, the numerical integration becomes computationally expensive.

\section{Physical interpretation}

\subsection{Resonances}

The recurring peaks seen in Fig.~\ref{fig:power} are reminiscent of the poles of the transmission coefficient in a quantum mechanical system that has bound states. It is indeed possible to interpret the observed behavior in a similar way.

The last term in \eqref{eqn:d2U} introduces a coupling between different modes, but also a self-coupling when $(\ell',m')=(\ell,m)$. The $\ell_*=0$ coefficient is $B_{\ell m;0}^{\ell m}=2$ for every $(\ell,m)$, as it simply describes the monopole, long-range component of the BH binary potential. We may then consider the $\ell_*>0$ terms as a small perturbation, and study the spectrum of the unperturbed equation,
\beq
\frac{\dd^2U_{\ell m}}{\dd r^2}=-2\mu EU_{\ell m}+\frac{\ell(\ell+1)}{r^2}U_{\ell m}-{}\begin{cases}
2GM\mu^2U_{\ell m}/R\,,\\
2GM\mu^2U_{\ell m}/r\,,
\end{cases}
\label{eqn:shell}
\eeq
where $E$ is the energy eigenvalue and the two cases correspond to $r<R$ and $r>R$ respectively. This potential is equivalent to that of a spherical shell of mass $M$ placed at $r=R$. At $r<R$, the solution is given by the regular spherical Bessel function of order $\ell$, while, at $r>R$, the regular solution is the second Whittaker function. The requirement that the two functions match smoothly at $r=R$ determines the eigenvalues $E=E_{n\ell}(\beta)$, with $n\ge\ell+1$. In the $\beta\to0$ limit, the radius of the shell vanishes and the hydrogenic spectrum is recovered, $E_{n\ell}(0)=-(GM)^2\mu^3/(2n^2)$.

In \eqref{eqn:d2U}, however, the energy is fixed by the condition \eqref{eqn:omega_m}. When $E_{n\ell}(\beta)=m\Omega$, we can thus expect that the $(\ell,m)$ mode will be resonantly excited. The same resonance condition has indeed been found in other systems where scalar bound states interact gravitationally with rotating binaries \cite{Baumann:2018vus,Baumann:2019ztm,Tomaselli:2024bdd,Tomaselli:2024dbw}. This condition can only be satisfied for $m<0$, i.e.\ for the quasi-bound modes, at some discrete values of $\beta=\beta_{n\ell m}$. These are shown as vertical lines in Fig.\ \ref{fig:power}, where they can be compared with the peaks observed in $P/\dot M_\slab{dm}$.

It is apparent that $\beta_{n2\,\sminus2}$ matches well the position of the main peaks, especially for $\ell_{\text{max}}=2$. We can thus loosely interpret the physics of the resonances as follows. For some specific values of $\beta$, the binary's frequency is just right to make the infalling DM waves transition into states that are gravitationally bound to the binary and co-rotate with it. These, however, are \emph{quasi}-bound, as the perturbation from the higher multipoles ($\ell_*>0$) introduces a nonzero decay width. Waves caught in these states are then eventually scattered away, and leave the system.

This qualitative interpretation is confirmed by Fig.~\ref{fig:profile}. Here, we compare the stationary DM configuration for two different values of $\beta$. Away from any of the resonances, most of the outgoing DM is in the $(0,0)$ mode, creating an approximately spherical standing wave density pattern. On resonance, instead, the $(2,-2)$ mode is significantly excited, and the DM configuration features nontrivial spiraling patterns, similar to those observed in numerical relativity simulations \cite{Bamber:2022pbs,Aurrekoetxea:2023jwk,Aurrekoetxea:2024cqd}.

Other modes besides $(2,-2)$ can be resonantly excited for large $\beta$. However, as is evident from Fig.~\ref{fig:power}, these resonances quickly become overlapping and crowded together, so that interpreting the complex features of $P/\dot M_\slab{dm}$ as due to individual resonances is difficult. One can notice, however, that some of the higher-mode resonances can introduce dips in $P/\dot M_\slab{dm}$ (rather than peak), such as the ones observed around $\beta_{54\,\sminus2}\sim11$ and $\beta_{64\,\sminus2}\sim14$.

\subsection{Perturbation theory}

For small enough $\beta$, where the DM scattering is less efficient, it should be possible to calculate the amplitude of the scattered waves as a small perturbation of the pure standing-wave scenario, where no scattering occurs. Let us restrict our attention to $\ell_{\text{max}}=2$, since at small $\beta$ that is a good approximation, as shown in Fig.~\ref{fig:power}. Among the modes with $\ell=2$, only $U_{22}$ has $k_m^2>0$, and it is thus the only mode that is able to carry energy to infinity. We thus need to compute $\abs{C_{22}}^2$, from which $P$ can be obtained via \eqref{eqn:power}. The equation of motion for $U_{22}$ can be seen as sourced by the dominant $U_{00}$ mode, through the quadrupole coupling $\ell_*=2$. If $\mathcal G(r,r')$ is the outgoing Green function, such that at large $r$ we have $\mathcal G(r,r')\approx A(r')e^{ik_2r}$ for some amplitude $A(r')$, we can thus approximate
\beq
\abs{C_{22}}^2\approx \abs{U_{22}}^2\approx\abs*{\int_0^\infty A(r')F_2(r')U_{00}(r')\dd r'}^2\,,
\label{eqn:perturbation}
\eeq
where the unperturbed $(0,0)$ mode is
\beq
U_{00}(r)\approx e^{-3i\pi/4}\sqrt{4\pi x}\,J_1(\sqrt{4x})\,,
\eeq
with $J_1$ the Bessel function of first order.

\begin{figure}
\centering
\includegraphics[]{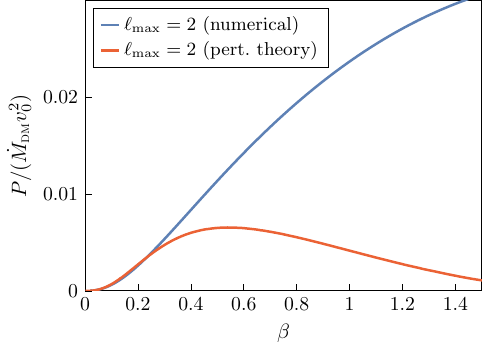}
\caption{Comparison of the numerical result \eqref{eqn:power} for $\ell_{\text{max}}=2$ with the result from perturbation theory, given in \eqref{eqn:perturbation}.}
\label{fig:perturbation}
\end{figure}

The result is shown in Fig.~\ref{fig:perturbation}. The perturbative result \eqref{eqn:perturbation} agrees well with the full numerical solution at $\beta\lesssim0.3$. This confirms that, for small $\beta$, the power is almost entirely carried to infinity by the quadrupolar $(2,2)$, whose amplitude depends only on the infalling DM flux, and not on all other angular momentum modes. The approximation however quickly degrades for larger $\beta$. In particular, neither the perturbative approach nor the resonance interpretation are able to satisfactorily explain the broad peak around $\beta\sim2$.

\subsection{Dynamical friction}

The energy lost to the scattering of DM can be interpreted as a form of dynamical friction \cite{Chandrasekhar:1943ys}, wherein the scattering of a medium off a moving object results in a drag force on the latter. This effect is often described in settings where the moving objects follow a straight trajectory through a uniform medium, and several works have recently quantified the friction force through an environment of wave DM \cite{Hui:2016ltb,Traykova:2021dua,Vicente:2022ivh,Traykova:2023qyv}. In that setting, the scattering of a plane wave off a BH results in a stationary density profile in the frame where the BH is at rest. Similarly, in our case the scattering off two orbiting BHs creates a stationary profile in the co-rotating frame. In both scenarios, this scattering backreacts on the orbit giving rise to a drag force.

Our calculation is also similar to works that studied dynamical friction on objects moving circularly through structures of wave DM, either self-gravitating \cite{Buehler:2022tmr,Annulli:2020lyc,Annulli:2020ilw,Duque:2023seg} or sustained by the gravity of another object, such as superradiant clouds \cite{Baumann:2021fkf,Baumann:2022pkl,Tomaselli:2023ysb,Brito:2023pyl}. By taking the limit where these DM structures become marginally bound, one should be able to recover the results of the present paper. We leave this analysis for a future work.

\section{Astrophysical implications}
\label{sec:astro}

\subsection{Estimating the mass flux}

In the previous sections, we described mathematically the scattering of wave DM off a BH binary. We now seek to make contact with real astrophysical settings and observations. The power emitted in scalar waves is directly proportional to the infalling mass flux $\dot M_\slab{dm}$. The larger $\dot M_\slab{dm}$, the stronger the backreaction of the DM scattering on the BH binary. A natural scenario to consider are thus the supermassive BH binaries found at galactic centers after galaxy mergers. This setting is particularly interesting because (a) the highest DM densities are found at the center of halos \cite{Hernquist:1990be,Navarro:1995iw,Navarro:2008kc,Seidel:1993zk,Schive:2014dra,Schive:2014hza,Veltmaat:2018dfz} (and they can be further enhanced by the supermassive BH's gravity \cite{Gondolo:1999ef,Ullio:2001fb,Gnedin:2003rj,Sadeghian:2013laa}), and (b) the supermassive BH's gravitational field can extend over a region that is larger than the DM's de Broglie wavelength.

Estimating $\dot M_\slab{dm}$ is not an easy task. This quantity depends significantly on the details of the environment around the BH binary, such as the velocity dispersion and the baryonic component, as well as on the DM mass $\mu$. Even though, in realistic settings, not all of the infalling DM is in the spherically symmetric $(0,0)$ mode, for simplicity we define $\dot M_\slab{dm}$ to only include the infalling waves with no angular momentum. This allows us to apply the results of \eqref{eqn:power} and Fig.~\ref{fig:power} directly to determine $P$. We will comment later on the accuracy of this approximation.

To gauge the possible range of values that $\dot M_\slab{dm}$ can take, we estimate it in two different ways. First, consider a spherically symmetric setup where, around the galactic center, the baryonic density $\rho_{\text{b}}$ is $\mathcal B$ times larger than that of DM $\rho$. Typical values of $\mathcal B$ are $\mathcal O(10)$ \cite{Linden:2014sra,McMillan:2011wd}. The binary's gravity dominates up to a distance $D$ such that $M\sim(4\pi/3)\rho_{\text{b}}D^3$, after which the gravity of the baryonic component becomes dominant. Similar to \cite{Hui:2019aqm,Hui:2022sri}, we may then imagine that DM falls in with zero mechanical energy from a distance $D$, in the $(0,0)$ mode towards the binary at the center. This implies $\dot M_\slab{dm}=4\pi D^2\rho\sqrt{2GM/D}$. Substituting the value of $D$, we find
\beq
\dot M_\slab{dm}^{(1)}=2M\sqrt{\frac{6\pi G\rho}{\mathcal B}}\,.
\label{eqn:dotM1}
\eeq
The estimate \eqref{eqn:dotM1} can be considered as optimistic, because it completely neglects the effect of velocity dispersion, assuming instead that all the DM falls in without angular momentum. Such a scenario could occur for small $\mu$, in halos that develop a solitonic core \cite{Seidel:1993zk,Schive:2014dra,Schive:2014hza,Veltmaat:2018dfz}.

One can instead consider a simplistic model where $\dot M_\slab{dm}$ is controlled by the velocity dispersion $v$: a plane wave $\psi=\sqrt{\rho/\mu}\,e^{i\mu vz}$. The ingoing $(0,0)$ component can be straightforwardly extracted with a partial wave expansion, leading to
\beq
\dot M_\slab{dm}^{(2)}=\frac{2\pi\rho}{\mu^2v}\,.
\label{eqn:dotM2}
\eeq
The estimate \eqref{eqn:dotM2} can be thought as conservative, because the resulting power $P$ neglects the contribution from modes with $\ell\ne0$. Indeed, as anticipated, in reality some of the infalling DM will not be in the $(0,0)$ mode. Waves with angular momentum number $\ell\ge1$, however, do not penetrate effectively their angular momentum barrier, which is placed at $r_\ell\sim\ell(\ell+1)/(2GM\mu^2)$. For large values of $\mu$, more waves with higher $\ell$ can come closer to the binary and contribute significantly to the scattering. The $(0,0)$ mode then becomes subdominant for large $\mu$, hence the $\dot M_\slab{dm}\propto\mu^{-2}$ scaling of \eqref{eqn:dotM2}.

When $r_\ell\gtrsim R$ (that is, $\beta\lesssim1$), only waves with $\ell=0$ can penetrate at small enough radii to resolve the binary and be efficiently scattered away. Therefore, for $\beta\lesssim1$, or
\beq
\mu\lesssim\SI{1.0e-20}{eV}\,\biggl(\frac{M}{10^9M_\odot}\biggr)^{-2/3}\biggl(\frac{f_\slab{gw}}{\SI{10}{nHz}}\biggr)^{1/3}\,,
\label{eqn:mu-limit}
\eeq
the entire mass flux $\dot M_\slab{dm}$ is indeed approximately in the $(0,0)$ mode. These are the values of $\mu$ where the results of Section~\ref{sec:scattering} can be safely applied.

\subsection{Gravitational-wave background}

The final stages of the evolution of a supermassive BH binary are driven by the emission of energy through GW radiation, which leads to the orbit shrinking. The pulsar timing residuals induced by a stochastic GW background resulting from a population of binaries evolving under GW radiation are predicted to depend on the GW frequency as $f_\slab{gw}^{-13/3}$ (see \cite{NANOGrav:2023gor} and references therein). When the binaries undergo some additional mechanism of energy loss with power $P$, such as environmental effects, their separation shrinks faster. As a consequence, the GW spectrum is rescaled proportionally to the reduced time they spend in each frequency interval: the timing residuals depend on the frequency as $f_\slab{gw}^{-13/3}/(1+P/P_\slab{gw})$, where $P_\slab{gw}=(2/5)\pi^{10/3}G^{7/3}(Mf_{\slab{gw}})^{10/3}$ is the power emitted in GWs. As GWs become increasingly efficient at higher frequencies, environmental effects are expected to produce a \emph{turnover}, making the spectrum flatter at low-frequencies. Data from the recent detection of a stochastic GW background show possible hints of a slope flatter than $-13/3$ \cite{NANOGrav:2023gor,NANOGrav:2024nmo}.

The scattering of DM discussed in the previous sections acts as an environmental effect on the GW background, similar to what has been studied in other works, such as~\cite{Ghoshal:2023fhh,Hu:2023oiu,Aghaie:2023lan,NANOGrav:2024nmo}. The estimates \eqref{eqn:dotM1} and \eqref{eqn:dotM2} for the mass flux allow us to quantify the order of magnitude of the effect. Employing the optimistic estimate \eqref{eqn:dotM1} to fix the value of $\dot M_\slab{dm}$, we can write the ratio $P/P_\slab{gw}$ as
\beq
\begin{split}
\frac{P^{(1)}}{P_\slab{gw}}={}&6.4\biggl(\frac{P/\dot M_\slab{dm}}{0.06v_0^2}\biggr)\biggl(\frac{\mathcal B}{10}\biggr)^{-1/2}\biggl(\frac{M}{10^9M_\odot}\biggr)^{-5/3}\\
&\times\biggl(\frac\rho{\SI{e+5}{GeV/cm^3}}\biggr)^{1/2}\biggl(\frac{f_\slab{gw}}{\SI{10}{nHz}}\biggr)^{-8/3}\,,
\end{split}
\label{eqn:PPgw1}
\eeq
while from the conservative estimate \eqref{eqn:dotM2} we get
\beq
\begin{split}
\frac{P^{(2)}}{P_\slab{gw}}={}&\num{1.8e-2}\biggl(\frac{P/\dot M_\slab{dm}}{0.06v_0^2}\biggr)\biggl(\frac{M}{10^9M_\odot}\biggr)^{-8/3}\\
&\times\biggl(\frac{\mu}{\SI{e-20}{eV}}\biggr)^{-2}\biggl(\frac{v}{\SI{400}{km/s}}\biggr)^{-1}\\
&\times\biggl(\frac\rho{\SI{e+5}{GeV/cm^3}}\biggr)
\biggl(\frac{f_\slab{gw}}{\SI{10}{nHz}}\biggr)^{-8/3}\,.
\end{split}
\label{eqn:PPgw2}
\eeq
Equations \eqref{eqn:PPgw1} and \eqref{eqn:PPgw2} have been written in such a way to isolate the ratio $P/(\dot M_\slab{dm}v_0^2)$, which can be directly read off equation \eqref{eqn:power} and Fig.~\ref{fig:power}. PTA data suggest that the GW background is dominated by BHs of masses around $10^9M_\odot$ \cite{NANOGrav:2023hfp}, so we choose this as a fiducial value for the binary's mass. The reference values for the dark matter density $\rho=\SI{e5}{GeV/cm^3}$ and velocity dispersion $v=\SI{400}{km/s}$ at galactic center have been chosen based on M87 data \cite{2009ApJ...700.1690G,DeLaurentis:2022nrv}, as it is one of the best-studied galaxies hosting a BH of this mass.

Due to the $M$-dependence in \eqref{eqn:PPgw1} and \eqref{eqn:PPgw2}, the resulting modification to the GW background depends on the mass function of supermassive BH binaries. This is however not the only information needed to obtain an accurate modelling of the signal: the efficiency of DM scattering will also depend on the mass ratio and on the orbital eccentricity, both of which are not included in our analysis. We thus find it more appropriate, at this stage, to postpone a more detailed study of the effect PTA spectrum to future work.

\begin{figure}
\centering
\includegraphics[]{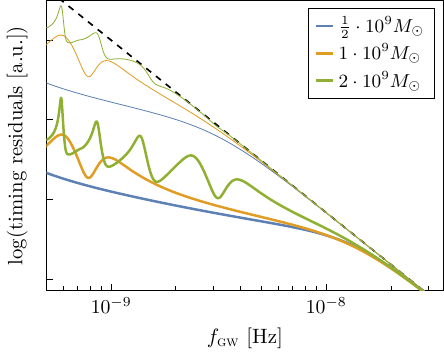}
\caption{Modification to the PTA timing residuals, in a simplistic model where the stochastic GW background is sourced by a population of BHs all having a mass of $0.5\cdot10^9M_\odot$ [{\color{Mathematica1}blue}], $1\cdot10^9M_\odot$ [{\color{Mathematica2}orange}], or $2\cdot10^9M_\odot$ [{\color{Mathematica3}green}], assuming $\mu=\SI{e-20}{eV}$. The curves show $f_\slab{gw}^{-13/3}/(1+P/P_\slab{gw})$ (with arbitrary normalization), where $P/P_\slab{gw}$ is given in \eqref{eqn:PPgw1} and \eqref{eqn:PPgw2}. The thick lines use the optimistic estimate \eqref{eqn:PPgw1} with $\mathcal B=10$, while the thin lines use instead the conservative estimate \eqref{eqn:PPgw2} with $v=\SI{400}{km/s}$. Both cases assume a DM density $\rho=\SI{e+5}{GeV/cm^3}$. The black dashed line denotes a power-law with $-13/3$ slope, i.e., the ``vacuum'' prediction.}
\label{fig:PTA-spectrum}
\end{figure}

Here instead, we illustrate the modification to the shape of the timing residuals in Fig.~\ref{fig:PTA-spectrum} by simplistically assuming a monochromatic mass function. Varying the value of $M$ allows us to have a qualitative understanding of how the results depend on it: higher masses shift the turnover to lower frequencies while bringing a larger number of resonances inside the band. The optimistic estimate \eqref{eqn:PPgw1} leads to a turnover around $\SI{10}{nHz}$ for $M=10^9M_\odot$, below which the spectrum flattens to a $f_\slab{gw}^{-5/3}$ slope. The resonant peaks of $P/\dot M_\slab{dm}$ observed in Fig.~\ref{fig:power} remain imprinted on the GW spectrum, raising the tantalizing possibility that information on the DM mass could be extracted from the spectrum data. However, it should be noted that the peaks' positions shift with $M$. We can thus expect a smooth mass function to at least partially smear out these features, while still preserving the $f_\slab{gw}^{-5/3}$ slope. The conservative estimate \eqref{eqn:PPgw2} can also produce observable effects around $\SI{1}{nHz}$, however the turnover is only pronounced when $M<10^9M_\odot$ or $\rho>\SI{e+5}{GeV/cm^3}$.

\section{Discussion}

\subsection{Particle dark matter}

The results presented in this paper have been derived within the model of wave DM, which reduces to that of standard non-interacting DM particles in the $\mu\to\infty$ limit. While the setup of our problem and the equations of motion hold for any value of $\mu$, the estimates we used in Sec.~\ref{sec:astro}, such as restricting the attention to the $(0,0)$ mode, break for large $\mu$. Furthermore, the values allowed by equation \eqref{eqn:mu-limit} are in tension with the most stringent constraints to date \cite{Rogers:2020ltq,Dalal:2022rmp}.

There are however suggestive hints that the particle DM limit is not less interesting. For instance, as can be seen in Fig.~\ref{fig:power}, the power $P/\dot M_\slab{dm}$ does not become smaller for large $\beta$. Instead, the height of the peaks seems to remain approximately constant, while their shape becomes more and more irregular as higher values of $\ell_{\text{max}}$ are used. For $\beta\to\infty$, one can expect the $\mu$-dependence to disappear and $P$ to asymptotically approach a finite limit. Similarly, it was found through numerical simulations in \cite{Aurrekoetxea:2023jwk} that the shortening of the merger time reaches a nonzero limit for large $\mu$.

Due to the computational challenge of using higher values of $\ell_{\text{max}}$ for large $\beta$, the particle limit should instead be dealt with by tracking the orbits of infalling test particles. This analysis would be particularly useful to determine whether the observable modification to the PTA spectrum should be generically expected in models of particle DM. We leave this question for a future work.

\subsection{\bf Outlook}

The results of this paper have been derived under a number of assumptions, most notably an equal-mass ratio, circular orbits and spherically symmetric boundary conditions for the infalling DM. A fully general treatment of the problem will require relaxing all these hypotheses. It will be interesting to see whether the features we observed, such as the resonant peaks in the power emitted, survive in more general cases. Such an improved modelling will also allow to derive a more accurate prediction for the modification of the PTA spectrum.

When discussing the astrophysical implications of DM scattering, in this paper we focused on the GW background. There may however exist other interesting consequences. For instance, the non-vanishing of the emitted power at large $\beta$ implies that the DM scattering is also efficient early in the evolution of a supermassive BH binary. This observation is particularly interesting, given the current uncertainty (known as the ``final parsec problem'' \cite{Begelman:1980vb,Milosavljevic:2002ht}) on which physical mechanisms can harden the binaries quickly enough to bring them to merger within astrophysical timescales. The possibility that DM, especially if ultralight, can help solve this problem has been discussed in a number of recent works \cite{Koo:2023gfm,Bromley:2023yfi,Alonso-Alvarez:2024gdz,Boey:2024dks}. Furthermore, the amplitude of the measured stochastic GW background had been found to be larger than theoretical predictions \cite{NANOGrav:2023hfp,Sato-Polito:2023gym,Sato-Polito:2024lew}, possibly indicating that the merging process is unexpectedly efficient. An early mechanism of energy loss might help address this puzzle.

The points mentioned above are only some of the many exciting connections between the physics of black holes, dark matter and extreme astrophysical environments. Through the influx of new data, gravitational-wave astronomy can be a key to shed light on their mysteries in the forthcoming years.


\section*{Acknowledgments}

G.M.T.\ is grateful to Lam Hui and Matias Zaldarriaga for helpful discussions, and to Andrea Caputo, Matias Zaldarriaga, and the authors of \cite{Bamber:2022pbs,Aurrekoetxea:2023jwk} for comments on the manuscript. G.M.T.\ acknowledges support from the Sivian Fund at the Institute for Advanced Study.

\appendix

\section{Stationary configuration}
\label{app:stationary}

As mentioned in the main text, the interpretation given in \cite{Bamber:2022pbs} of the universal density profile is that it is \emph{quasi}-stationary, its amplitude growing over time as more dark matter falls in. It is not easy to conclusively determine from the simulations whether this trend continues indefinitely or is a transient towards a truly stationary regime.

We present here a proof that any universal density profile that co-rotates with the binary must be stationary. For this purpose, it is convenient to switch to the co-rotating frame, where the acceleration $\vec g$ of a fluid element receives contributions from the centrifugal and Coriolis apparent forces,
\beq
\vec g=-\vec\Omega\times(\vec\Omega\times\vec r)-2\vec\Omega\times\vec v-\vec\nabla V\,.
\eeq
Furthermore, we use the Madelung formulation, where the Schrödinger equation \eqref{eqn:schrodinger} is recast into a pair of hydrodynamical equations for the variables
\beq
\rho=\mu\abs{\psi}^2,\qquad\vec v=\frac{i}{2\mu\abs{\psi}^2}(\psi\vec\nabla\psi^*-\psi^*\vec\nabla\psi)\,.
\eeq
The equations take the form of mass conservation,
\beq
\frac{\partial\rho}{\partial t}+\vec\nabla\cdot(\rho\vec v)=0\,,
\label{eqn:continuity}
\eeq
and the Euler equation,
\beq
\frac{\partial\vec v}{\partial t}+(\vec v\cdot\vec\nabla)\vec v=\vec g+\frac1{2\mu^2}\vec\nabla\biggl(\frac{\nabla^2\sqrt\rho}{\sqrt\rho}\biggr)\,.
\label{eqn:euler}
\eeq

A rigidly co-rotating fluid configuration, which remains self-similar while its amplitude varies with time, can be described in general as
\beq
\rho(t,\vec r)=D(t)\rho_0(\vec r),\qquad \vec v(t,\vec r)=S(t)\vec v_0(\vec r)\,,
\label{eqn:factorization}
\eeq
for some density field $\rho_0$ and velocity field $\vec v_0$, where the functions $D(t)$ and $S(t)$ dictate the time variation. Substituting \eqref{eqn:factorization} into \eqref{eqn:continuity}, we obtain
\beq
\frac{\dot D}{DS}=-\frac{\vec\nabla\cdot(\rho_0\vec v_0)}{\rho_0}\equiv\Lambda\,,
\label{eqn:DS-continuity}
\eeq
where $\Lambda$ must be a constant, because the LHS only depends on time, while the RHS only depends on $\vec r$.

Let us consider equation \eqref{eqn:euler} along the $z$ axis. By $\mathbb{Z}_2$ rotational symmetry, the velocity must be $z$-directed. The $z$ component of \eqref{eqn:euler} then reads
\beq
\dot S+S^2\partial_zv_{0z}=\frac1{v_{0z}}\biggl(g_z+\frac1{2\mu^2}\partial_z\biggl(\frac{\nabla^2\sqrt{\rho_0}}{\sqrt{\rho_0}}\biggr)\biggr)\,.
\eeq
Neither the centrifugal nor the Coriolis forces contribute to $g_z$ along the $z$ axis. At any $z$, the RHS is thus time-independent, and so must be the LHS: $\ddot S+2\dot SS\partial_zv_{0z}=0$. Evaluating this equation at two points on the $z$ axis where $\partial_zv_{0z}$ takes different values, we find $\ddot S=\dot SS=0$, hence $S$ must be a constant.

Because $S$ is a constant, from \eqref{eqn:DS-continuity} we conclude that $D$ must depend exponentially on time, $D\propto e^{\Lambda St}$. A profile whose density grows exponentially with time can certainly not be fed by a \emph{constant} influx of dark matter from infinity. On the other hand, a decaying exponential implies the quick disappearance of the density profile. We conclude that it must be $\Lambda=0$, i.e., the co-rotating dark matter profile must be stationary.

\clearpage
\bibliography{main}

\end{document}